\documentclass[a4paper,fleqn,usenatbib,useAMS]{mnras}

\usepackage{graphicx}	% Including figure files
\usepackage{amsmath}	% Advanced maths commands
\usepackage{amssymb}	% Extra maths symbols
\usepackage{multicol}        % Multi-column entries in tables
\usepackage{bm}		% Bold maths symbols, including upright Greek
\usepackage{pdflscape}	% Landscape pages
\newcommand{\vect}[1]{\mathbfit{#1}}  %Vector font for roman character
\newcommand{\mat}[1]{\mathbfss{#1}}   % Matrix font for roman character
\newcommand{\field}{\bm{\delta}}
\newcommand{\dd}{\mathrm{d}}
\usepackage[T1]{fontenc}
\usepackage{ae,aecompl}

\usepackage{newtxtext,newtxmath}
\usepackage{soul}
\title[Quadratic genetic modifications]
{Quadratic genetic modifications: a streamlined route to \\
cosmological simulations with controlled merger history}

\author[M. P. Rey and A. Pontzen]
{Martin P. Rey$^{1}$\thanks{Contact e-mail: \href{martin.rey.16@ucl.ac.uk}{martin.rey.16@ucl.ac.uk}},
{Andrew Pontzen$^{1}$}
\\
$^{1}$Department of Physics and Astronomy, University College London, London WC1E 6BT, UK}

\date{Draft Version, 14/06/2017}

\pubyear{2017}

\begin{document}
\label{firstpage}
\pagerange{\pageref{firstpage}--\pageref{lastpage}}
\maketitle

\begin{abstract}
  Recent work has studied the interplay between a galaxy's history and
  its observable properties using ``genetically modified''
  cosmological zoom simulations. The approach systematically
  generates alternative histories for a halo, while keeping its
  cosmological environment fixed.
  Applications to date altered linear properties of the
  initial conditions such as the mean overdensity of specified regions; we
  extend the formulation to include quadratic features such as
  local variance, which determines the overall importance of smooth
  accretion relative to mergers in a galaxy's history.
  We introduce an efficient algorithm for this new class of modification
  and demonstrate its ability to control the variance of a
  region in a one-dimensional toy model. Outcomes of this
  work are two-fold: (i) a clarification of the formulation of genetic
  modifications and (ii) a proof of concept for quadratic modifications
  leading the way to a forthcoming implementation in cosmological
  simulations.
\end{abstract}


%%%%%%%%%%%%%%%%%%%%%%%%%%%%%%%%%%%%%%%%%%%%%%%%%%

%%%%%%%%%%%%%%%%% BODY OF PAPER %%%%%%%%%%%%%%%%%%

\section{Introduction} \label{sec:intro}

\defcitealias{HR1991}{HR91}

Mergers and accretion are thought to play a key role in shaping the
observed galaxy population; in the prevailing cosmological
paradigm merger histories are in turn seeded from random inflationary
perturbations. Numerical studies must make
inferences about the galaxy population from a finite sample of such
histories. Due to the limited computer time available, this generates
a tension between resolution (for resolving the
interstellar medium) and volume (for adequately sampling histories).

One attempt to sidestep this problem is to create and study a small
number of carefully controlled tests of the relationship between a
galaxy's history and its observable properties.  This has long
been attempted in idealised, non-cosmological settings
(e.g. \citealt{Hernquist1993, DiMatteo2005, Hopkins2012}).
More recently, \cite{Roth2016} proposed performing such tests within a fully cosmological
environment by constructing a series of closely-related initial
conditions with targeted ``genetic modifications'' (hereafter
GMs). The formalism resembles that of constrained
realisations (\citealt{BBKS, Bertschinger1987, HR1991}, hereafter
\citetalias{HR1991}) which generates realisations of Gaussian
random fields satisfying user-defined constraints on
initial densities, velocities or potentials (e.g.
\citealt{Bertschinger2001}). Simulations based on constrained realisations have been extensively
applied to recreating the local universe using observed galaxy
distributions as constraints (for recent examples see \citealt{HeB2013, Wang2016, Sorce2016a, Hoffman2017}).

Despite a resemblance, genetically modified simulations are markedly
different from constrained simulations. The process of GM involves
creating multiple versions of the initial conditions, each with
carefully selected small changes. By re-simulating each scenario it
becomes possible to study how the changes affect the non-linear
evolution of structure. For example, modifications can be chosen such
that they enhance or suppress merger ratios in incremental steps and
so vary a galaxy's history in a systematic and controlled way. The
first application of this technique in a hydrodynamic simulation was
made by \citet{Pontzen2017}; that work focuses on the response of a
galaxy's central black hole and its ability to quench star formation
as the merger history is changed gradually.  Unlike studies based on
fully idealised merger simulations, the GM-based approach is able to
capture the effects of gradual gas accretion from filaments which is
essential when probing the balance between star formation and black
hole feedback.

On a technical level, \citet{Pontzen2017} used multiple linear
modifications to alter the merger history. Such a method requires
human effort on two fronts: (i) to identify and track particles
forming the merging substructures; and (ii) to tune the modifications
and understand their effects on one another. For instance, GMs
suppressing a merger tend to increase the mass of other nearby
substructures, which complicates interpretation of the final results
(see section 2.3 and figure 2 of \citealt{Pontzen2017}). Bypassing
this behaviour would be possible by individually identifying all
substructures and  demanding the algorithm fix each
one. However, the spiralling complexity of the setup makes this option
unattractive.

Another possibility, which is the primary aim of the present paper, is
to find a new type of modification which automatically suppresses the
merger ratios of \textit{all} large substructures in a target galaxy's
history. Such a modification would smooth the expected history while
keeping its final mass and overall environment fixed.
These modifications must be applicable to cosmological simulations,
so our objective is an algorithm that remains tractable even working
with fields on multidimensional grids. To achieve this
goal, we start by
clarifying the formulation of GMs (Section \ref{sec:reformulation}).
We then expand the framework to quadratic modifications
(Section~\ref{sec:quadratic}), allowing control over the variance at
different scales to tackle the problem of multiple mergers. We
demonstrate the feasibility of our method on a one-dimensional model
(Section~\ref{sec:demonstration}); in forthcoming work we will
demonstrate the implementation for a full 3D zoom simulation. Results
are discussed in Section \ref{sec:discussion} and we conclude in
Section~\ref{sec:concl}.

% Linear modifications section
%%%%%%%%%
\section{Linear constraints and modified fields} \label{sec:reformulation}
In this Section, we contrast the method of constraints
\citepalias{HR1991} against that of linear genetic modifications. The aim is
to clarify the status of the latter as a building block for non-linear
GMs, which are introduced in Section~\ref{sec:quadratic}.

\subsection{Constrained ensemble}   \label{subsec:sampling_linear}

\begin{figure}
  \centering
    \includegraphics[width=\columnwidth]{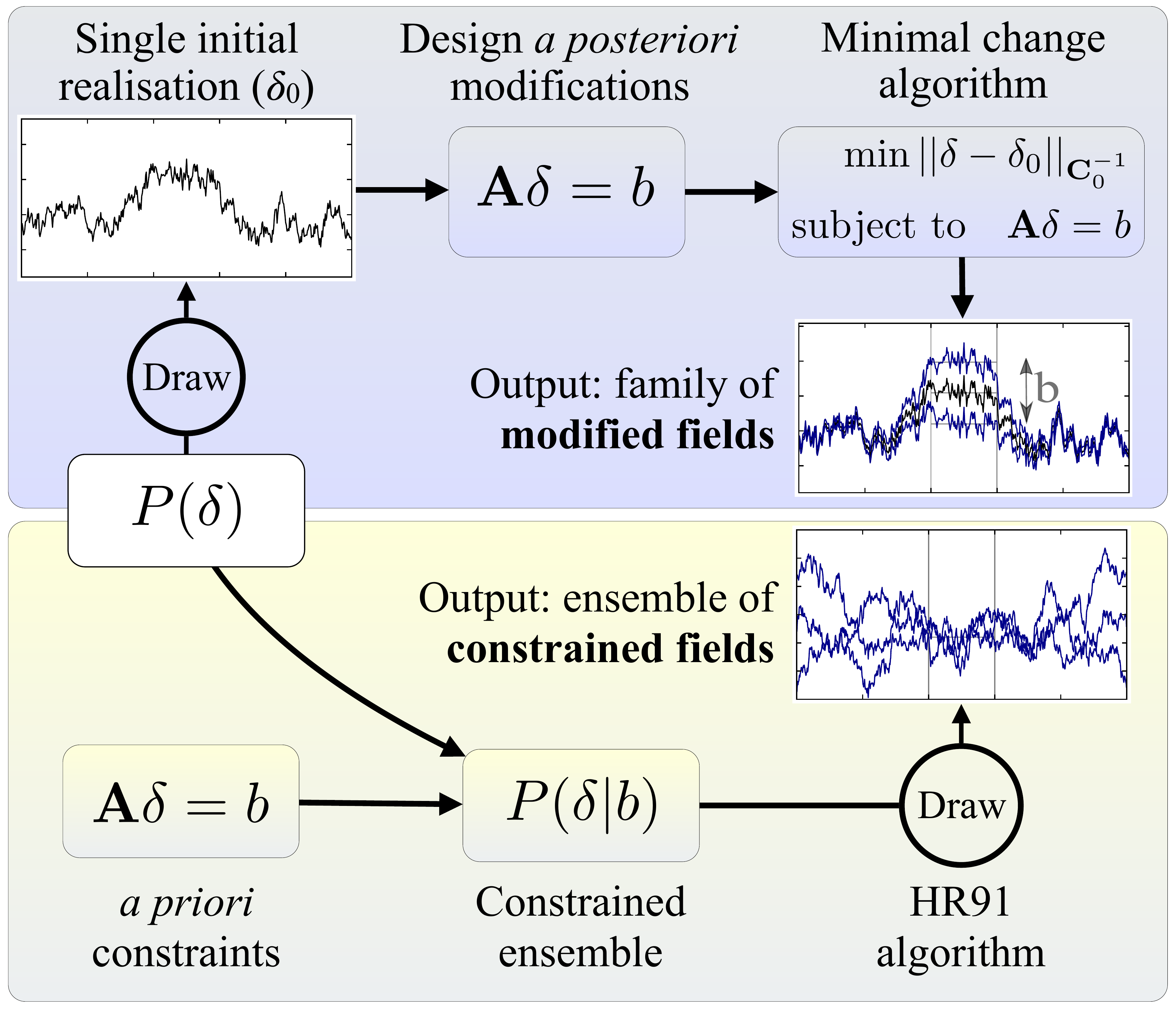}

    \caption{GMs and constrained ensembles are
      two techniques to generate targeted initial conditions
      for numerical simulations. They have markedly different
      motivations and properties despite sharing similar
      mathematics. We illustrate the differences by showing the flow
      of information in the two cases. \textit{Upper Panel:} In the GM
      case, a single initial realisation (black) is first drawn from
      the underlying ensemble.  Next, modifications are designed to
      alter chosen properties of this realisation;
      each modification therefore depends on the specific $\field_0$. The
      modified fields are computed by demanding minimal changes while
      satisfying the requested modifications.  In the illustrated
      example, we create two modified fields with enhanced and reduced
      mean values, corresponding to two different values of $\vect{b}$
      inside the target region. \textit{Lower panel:} In the
      constrained ensemble case, the constraints are independent of
      any particular realisation and are used to define the ensemble
      $P(\field|b)$.  This ensemble is efficiently sampled using the
      \citetalias{HR1991} technique.  In this example, three fields
      are drawn and by construction satisfy the same mean value inside
      the target region.}
 \label{fig:overview}
\end{figure}

We start by reviewing the construction of
constrained ensembles (see bottom panel of Figure~\ref{fig:overview}).
In this case, constraints must be known \textit{a priori},
i.e. independently of any specific realisation.
Constrained ensembles are therefore particularly
useful when using observations as external inputs to constrain
numerical simulations.

Consider a Gaussian random field sampled at $n$ points to create
a vector $\field$
with covariance matrix $\mat{C}_0= \langle \, \field \,
\field^{\dagger} \, \rangle$.
The \citetalias{HR1991} algorithm allows for an arbitrary number
(denoted $p$) of
linear constraints to be placed on $\field$; these can be expressed as
$\mat{A} \, \field= \vect{b}$
where $\mat{A}$ is a $p \times n$ matrix and $\vect{b}$ is a length-$p$ vector.

We start by constructing the ensemble of all fields $\field$ satisfying the constraint for a chosen $\vect{b}$, i.e.
$P(\field|\vect{b})$. Applying Bayes' theorem, the probability reads
\begin{equation}
\label{eq:likelihood}
P(\field | \vect{b}) = \frac{P(\vect{b}| \field) \, P(\field)}{P(\vect{b})} \, .
\end{equation}
Using the fact that $P(\field)$ is Gaussian and disregarding normalization,
this relation becomes
\begin{equation}
P(\field | \vect{b}) \propto
              \delta_D(\mat{A} \, \field - \vect{b}) \,
              \exp(-\frac{1}{2} \field^{\dagger} \mat{C}_0^{-1} \field)
              \, ,\label{eq:constrained_stats_with_delta}
\end{equation}
where $\delta_D$ is the ($p$-dimensional) Dirac delta function.

This expression suggests a brute force sampling solution: we could
draw many trial $\field$s from the original ensemble
and keep only the ones satisfying the constraints (within some tolerance). This solution
is, however, computationally inefficient. Making use of the fact
that the Dirac delta function can be represented as the zero-variance
limit of a Gaussian, we can instead derive the following results (\citealt{Bertschinger1987}):
\begin{align}
\label{eq:constrained_stats}
&P(\field | \vect{b}) \propto
      e^{-\frac{1}{2} (\field - \bar{\field})^{\dagger} \mat{C}^{-1} (\field - \bar{\field})}
       \, , \ \text{with} \nonumber  \\
&\bar{\field}(\vect{b}) =
      \mat{C}_0 \, \mat{A}^{\dagger} \, (\mat{A} \, \mat{C}_0 \, \mat{A}^{\dagger})^{-1} \, \vect{b}
      \, , \ \text{and} \nonumber \\
&\mat{C} = \mat{C}_0 - \mat{C}_0 \, \mat{A}^{\dagger}
          (\mat{A} \, \mat{C}_0 \, \mat{A}^{\dagger})^{-1} \, \mat{A}\, \mat{C}_0 \, ,
\end{align}
where $\bar{\field}$ and $\mat{C}$ are the expectation and the covariance of the
Gaussian distribution $P(\field | \vect{b})$. By construction, all
fields drawn from this distribution will satisfy the constraints
($\mat{A}\field = \vect{b}$).

\citet{HR1991} pointed out a convenient shortcut for efficiently
sampling from the distribution specified by Equation~\eqref{eq:constrained_stats}. Starting from a draw of the unconstrained ensemble, $\field_0$,
we calculate $\vect{b}_0 = \mat{A} \, \field_0$. One can then rewrite
$\field_0$ as the sum of the mean field $\bar{\field}(\vect{b}_0)$ from Equation
(\ref{eq:constrained_stats}) and a residual term
$\field_{\mathrm{residual}}$, defined by:
\begin{align}
\field_{\text{residual}}  & \equiv \field_0 - \bar{\field}(\vect{b}_0) \nonumber \\
& = \field_{0} - \mat{C}_0 \, \mat{A}^{\dagger} \,
          (\mat{A} \, \mat{C}_0  \, \mat{A}^{\dagger})^{-1} \, \vect{b}_0 \, .
\end{align}
From here, a draw from the constrained ensemble $\field_1$
can be generated by recombining the residuals with the corrected mean
$\bar{\field}(\vect{b})$:
\vspace{-0.3cm}
\begin{equation}
\field_1 = \mat{C}_0 \, \mat{A}^{\dagger} \,
            (\mat{A} \, \mat{C}_0  \, \mat{A}^{\dagger})^{-1} \, \vect{b} + \field_{\text{residual}} \, .
\end{equation}
To verify this procedure draws samples $\field_1$ from the constrained
distribution,  one first writes the mapping from $\field_0$
to $\field_1$ in a single step:
\begin{equation}
\field_1 = \field_0 -
          \, \mat{C}_0 \, \mat{A}^{\dagger} \,
          (\mat{A} \, \mat{C}_0  \, \mat{A}^{\dagger})^{-1} \, (\mat{A} \, \field_0 - \vect{b} ) \, .\label{eq:HR91-modification-step}
\end{equation}
Then, by calculating $\langle \field_1 \rangle$ and
$\langle \field_1 \field_1^{\dagger} \rangle$, it is possible to
check that the ensemble has the correct mean and covariance from
Equation (\ref{eq:constrained_stats}). The fact that $\field_1$ is
Gaussian follows from its construction as a linear transformation of
$\field_0$.  The underlying efficiency of this method is that the
covariance matrix in Equation (\ref{eq:constrained_stats}), does not
depend on the value of $\vect{b}$, allowing the
$\field_{\mathrm{residual}}$ term to be the same for both
expressions.

In summary, the \citetalias{HR1991} algorithm creates a draw from the constrained ensemble
in two steps, using the realisation $\field_0$ as an intermediate construction tool.
It provides a computationally efficient way of generating Gaussian
constrained fields.

\subsection{Genetic modifications} \label{subsec:optimization_linear}
We now turn to GMs (see upper panel of Figure~\ref{fig:overview}) to
constrast their formulation with that of constrained fields.
The GM procedure can be summarized as follows:
\begin{enumerate}
	\item Draw the unmodified realisation $\field_0$.
	\item Define the modifications by choosing which
          properties of $\field_0$ are to be modified. Unlike in the
          constrained field case, this is accomplished with
          reference to specific features of the $\field_0$
          realisation (e.g. the location and properties of particular haloes).
          This reflects how GMs are intended
          for constructing numerical experiments
          rather than for recreating observationally motivated scenarios. We
          focus first on linear modifications, i.e.  of the form
          $\mat{A} \, \field = \vect{b}$.
	\item Create the modified field (or multiple modified
          fields with different values of $\vect{b}$). We require
          changes between fields to be as small as possible, which relies on the
          definition of a distance in field space.
          In the context of Gaussian fields, the only available metric is
          defined by the $\chi^2$ distance,
          \begin{equation}
            \chi^2 \equiv ||\, \field \,||_{\mat{C}_0^{-1}}^2=\field^{\dagger} \,
            \mat{C}_0^{-1} \, \field \, .
          \end{equation}
\end{enumerate}
Consequently, GMs can be formulated as finding the modified field
solution of the following optimization problem:
\begin{gather}
\label{eq:optim_linear}
\begin{split}
\qquad &\min_{\field} \quad ||\, \field-\field_0 \,||_{\mat{C}_0^{-1}}^2 \, , \\
\qquad &\text{subject to} \quad \mat{A} \, \field = \vect{b} \, .
\end{split}
\end{gather}
The problem is solved by minimising the Lagrangian
\begin{equation}
\mathcal{L} \equiv (\field-\field_0)^{\dagger}\, \mat{C}_0^{-1} \, (\field-\field_0) +
                \bm{\lambda}^{\dagger} \, (\mat{A}\, \field - \vect{b}) \, ,
\end{equation}
where $\bm{\lambda}$ is a vector of size $p$ containing the Lagrange multipliers for each modification.

By differentiating to find critical points
with respect to $\field$ and $\bm{\lambda}$, we
obtain a system of two vector equations with the solution
\begin{equation}
\label{eq:optimum_linear}
\field_1  = \field_0 -
            \, \mat{C}_0 \, \mat{A}^{\dagger} \, (\mat{A} \, \mat{C}_0  \, \mat{A}^{\dagger})^{-1}
            \, (\mat{A} \, \field_0 - \vect{b} ) \, ,
\end{equation}
where $\field_1$ is the modified field.

Equation~\eqref{eq:optimum_linear} has regenerated Equation~\eqref{eq:HR91-modification-step}
using a different motivation and
derivation. To summarise:
\begin{itemize}
\item In the case of~\eqref{eq:HR91-modification-step},
$\field_0$ is an intermediate construct that is never used in a
simulation; it only exists to aid finding $\field_1$, which is
a sample from the distribution~\eqref{eq:constrained_stats}.

\item In the case of~\eqref{eq:optimum_linear}, $\field_0$ and
  $\field_1$ are put on equal footing. They are both initial condition
  fields drawn from the original, underlying ensemble $P(\field)$.
  The fact that the modifications (choice of $\mat{A}$ and $\vect{b}$)
  depend on $\field_0$, as emphasised by \cite{Porciani2016}, does not
  impact this interpretation.

\item We show in Appendix~\ref{app:sampling_quadratic}
that in the case of non-linear constraints, there is no joint
expression for GMs and \citetalias{HR1991}, formalising their intrinsic
difference.
\end{itemize}

GMs should therefore be seen as a mapping between fields of the same ensemble.
A family of modified fields is
generated by choosing multiple values for $\vect{b}$; the resulting
mapping between members of the family is continuous and
invertible. These properties are highly valuable for providing
controlled tests, allowing for systematic exploration of the effects of formation
history on a galaxy.

While the algorithm makes the minimal changes to the field,
$\field_1$ may still not be a
particularly likely draw from $P(\field)$ if the modifications are too
extreme. To quantify the level of alteration, the relative likelihood
of the two fields is given by $\exp{(- \Delta \chi^2/2)}$ with
\begin{equation}
  \Delta \chi^2 = \field_1^{\dagger} \, \mat{C}_0^{-1} \, \field_1  -
                  \field_0^{\dagger} \, \mat{C}_0^{-1} \, \field_0 \, .
\end{equation}
As long as $\Delta \chi^2$ stays small,
 we can regard the modified and unmodified
fields as similarly likely draws from $\Lambda \text{CDM}$ initial
conditions.

Turning $\Delta \chi^2$ into a precise quantitative statement about
the relative abundance of a particular galactic history remains a
topic for future research. It relies on knowing the detailed
Jacobian relating the initial conditions to properties of the
final galaxy. This can so far only be estimated, and only in simple
scenarios such as small modifications to the halo mass
\citep{Roth2016}. There are multiple possible
modifications (i.e. choices of $\mat{A}$ and $\vect{b}$) leading to a given
effect in the target galaxy history \citep{Porciani2016}; some will carry a
smaller $\Delta \chi^2$ cost than others. Finding the minimum-cost
route to a given change in the non-linear universe is not the aim of GMs;
to perform galaxy formation experiments,
we only need to find one choice of modification
with an \textit{acceptably} small $\Delta \chi^2$ penalty.

%%%%%%%%%
% Quadratic modifications section
%%%%%%%%%

\section{Extension to quadratic modifications} \label{sec:quadratic}

The main aim of this paper is to formulate modifications that control
the variance of a field. The variance on
scales smaller than the parent halo scale relates to the
number of substructures in haloes (\citealt{Press1974, Bond1991}), and
is therefore a proxy for the overall importance of mergers.
It is important to distinguish variance modifications of a region from alterations
to the power spectrum. The power spectrum defines only the average variance over the entire box,
and over all possible realisations. We propose on the other hand to modify the local variance, targeting
\textit{one} region of interest and making minimal changes to the
remaining structures. Another way to picture this goal is as follows. In any one
stochastic ensemble, two realisations might by chance have enhanced or
reduced variance in an area. Our procedure aims to map between such realisations
rather than to modify the underlying power spectrum.

Variance is quadratic in the field value and therefore the approach in
Section~\ref{subsec:optimization_linear} cannot be applied directly.
One natural formulation of the problem is through a new minimisation
problem (analogous to the original linear case):
\begin{gather}
\label{eq:optim_quadratic}
\begin{split}
&\qquad \min_{\field} \quad ||\, \field-\field_0 \,||_{\mat{C}_0^{-1}}^2 \, , \\
&\qquad\text{subject to} \quad \field^{\dagger} \, \mat{Q} \, \field= q \, ,
\end{split}
\end{gather}
where $\mat{Q}$ is a $n \times n$ matrix and $q$ is a scalar.
We can assume without loss of generality that $\mat{Q}$ is Hermitian.
For a suitable choice of $\mat{Q}$ (see Section~\ref{sec:demonstration}), $q$ specifies the
variance of a chosen region.

Following a similar approach to the linear modifications, we introduce
the Lagrangian
\begin{equation}
\mathcal{L} = (\field-\field_0)^{\dagger}\, \mat{C}_0^{-1} \, (\field-\field_0) +
              \mu \, (\field^{\dagger} \, \mat{Q} \, \field - q) \, ,
\end{equation}
where $\mu$ is a scalar Lagrange multiplier associated with the quadratic
modification.
Searching for critical points, we obtain two equations relating the
modified field $\field_1$ and the multiplier:
\begin{align}
&\field_1=(\mat{I}+\mu \, \mat{C}_0 \, \mat{Q})^{-1} \, \field_0 \, , \ \text{and}
\label{eq:field1-from-mu}\\
&\field_0^{\dagger} \, (\mat{I}+\mu \, \mat{C}_0 \, \mat{Q})^{-1} \,
			\mat{Q} \, (\mat{I}+\mu \, \mat{C}_0 \, \mat{Q})^{-1} \, \field_0 = q
\label{eq:mu-to-q} \, .
\end{align}
Equation~\eqref{eq:field1-from-mu} and~\eqref{eq:mu-to-q}
provide a closed system for $\mu$ and $\field_1$ given a target $q$. Unlike
the linear case, the system can not be solved analytically. A possibility
would be to solve Equation~\eqref{eq:mu-to-q} numerically for $\mu$ but direct matrix
inversions are prohibited due to their computational cost.
One would therefore need to perform approximate matrix inversion at each step of a root-finding scheme
for $\mu$, making the worst-case complexity of such method infeasible.

There are moreover deeper reasons why such procedures can not be
straightforwardly adapted to GMs.
In the linear case discussed above, we defined GMs as a continuous and
invertible mapping. Both of these properties are lost
when looking at Equations~\eqref{eq:field1-from-mu}
and~\eqref{eq:mu-to-q}. First, it is not clear that Equation~\eqref{eq:mu-to-q} has a real
  solution for $\mu$. Consequently  a real-valued $\field_1$ may not exist\footnote{We
    note in passing that, since variance is a positive quantity,
    $\mat{Q}$ is a positive semi-definite matrix. By definition,
    $\mat{C}_0$ is positive definite. These conditions ensure
    that $\field_1$ is unique if it exists -- but they do not
    guarantee existence.} for any chosen value of $q$.

Second, the relationship between $\field_0$ and $\field_1$ is
  asymmetric: if a new field $\field'$ is constructed by taking $q$
  back to its original value $q_0$, we will have
  \begin{equation}
  \field' = \left( \mat{I} + (\mu+\mu') \mat{C}_0\mat{Q} + \mu \mu' \left(\mat{C}_0\mat{Q}\right)^2\right)^{-1}
  \field_0 \, \textrm{,}
  \end{equation}
  for suitable choices of $\mu$ and $\mu'$. To obtain a solution
  $\mu'$ allowing recovery of the initial field
  ($\field_0 = \field'$), it must hold that
  $\mat{C}_0 \mat{Q} \propto (\mat{C}_0 \mat{Q})^2$. This will not
  generally be the case for our applications, and so we conclude that
  in general $\field' \ne \field_0$. Such asymmetry would be
  problematic for GM; the sense of a unique `family' of fields is
  lost.

The combination of computational intractability
and loss of key properties for GMs lead us to focus on an alternate method.
We describe next a Newton-like method which efficiently approximates a solution to the optimization problem,
Equation~\eqref{eq:optim_quadratic}, while reinstating the desired
properties of the GM mapping.

\subsection{Linearised solution} \label{subsec:linearised method}

In this section, we restate the quadratic problem in a way that has a
guaranteed solution and that generates a single family as a function
of $q$. The trick is to make only infinitesimally small changes to the
value of $q$, building up finite changes by following a path through
field space that is \textit{locally} minimal. This leads to an
iterative procedure for quadratic genetic modifications, which we will
demonstrate is both unique and computationally tractable.

\subsubsection{One infinitesimal step} \label{subsubsec:step} We start
by defining the displacement $\bm{\epsilon}$ from the unmodified field
$\field=\field_0 + \bm{\epsilon}$; for sufficiently small changes we
may then neglect
$\mathcal{O}(\bm{\epsilon}^2)$ terms. We
will discuss in Section~\ref{subsubsec:nsteps} how to practically
decompose a macroscopic change into a series of such minor
modifications.

At first order, the updated variance (or other quadratic property) is given by
\begin{equation}
\label{eq:inf_modification}
\field^{\dagger} \, \mat{Q} \, \field =
\field_0^{\dagger} \, \mat{Q} \, \field_0 +
2 \, \field_0^{\dagger} \, \mat{Q} \, \bm{\epsilon} +
\mathcal{O}(\bm{\epsilon}^2) \, ,
\end{equation}
where we have assumed $\field$ is real and made use of the previously stated Hermitian assumption, $\mat{Q}^{\dagger} = \mat{Q}$.
Having linearised the modification, we can now find
an analytic solution for the displacement and the multiplier $\mu$:
\begin{align}
\label{eq:epsilon_definition}
&\bm{\epsilon}=\mu \, \mat{C}_0 \, \mat{Q} \, \field_0 \, , \ \text{with}\\
&\mu = \frac{1}{2} \, \frac{q - \field_0^{\dagger} \, \mat{Q} \, \field_0}
{\field_0^{\dagger} \, \mat{Q} \, \mat{C}_0 \, \mat{Q} \, \field_0} \,
  . \label{eq:mu-for-infinitessimal}
\end{align}
Equation~\eqref{eq:mu-for-infinitessimal} does not involve matrix inversions
and can therefore be efficiently evaluated, even in a 3D cosmological simulation context.

\subsubsection{Building finite changes by successive infinitesimal updates} \label{subsubsec:algo}
We now want to construct a macroscopic change in the field by iterating
the infinitesimal steps of Equation~\eqref{eq:epsilon_definition}.
 Performing a finite number of steps $N$, the modified field
reads:
\begin{equation}
  \field_1 = \prod_{j=0}^{N} \left( \mat{I} + \mu_j \,
    \mat{C}_0 \, \mat{Q} \right) \field_0 \, ,\label{eq:finite-N-steps}
\end{equation}
where $\mu_j$ is the Lagrange multiplier at step $j$. The value of each $\mu_j$ depends
on how the fixed interval is divided, i.e. implicitly on $N$. In the
limit of increasing number of steps, each individual
$\mu_j$ becomes infinitesimally small and the final
solution is
\begin{align}
\field_1 &= \lim_{\substack{N\to\infty \\ \mu_j \to 0}}\,\, \prod_{j=0}^{N} \left( \mat{I} + \mu_j \,
  \mat{C}_0 \, \mat{Q} \right) \field_0 \nonumber \\
  & =
  \prod_{j=0}^{\infty} \exp \left( \mu_j \mat{C}_0 \, \mat{Q} \right)
   \equiv \exp \left( \alpha \mat{C}_0 \, \mat{Q} \right) \field_0 \, ,
  \label{eq:matrix-exp-solution}
\end{align}
where $\alpha = \sum_{j=0}^{\infty} \mu_j$ is
the overall displacement and is finite.
The right-hand side of
Equation~\eqref{eq:matrix-exp-solution} defines the matrix exponential
operator, which is guaranteed to exist and is invertible.

The matrix exponential is a useful formal expression to show that
there is a unique result, but does not help computationally since
the required value of $\alpha$ to reach the objective
$\field_1^{\dagger} \mat{Q}\, \field_1=q$  is unknown. In
practice, we use the finite approximation, Equation~\eqref{eq:finite-N-steps}.
The $\mu_j$ at each step are chosen by
targeting $N$ intermediate modifications linearly spaced between the
starting value $q_0\equiv\field_0^{\dagger} \mat{Q}\, \field_0$ and
the target $q$. At each step, $\mu_j$ is calculated using
Equation~\eqref{eq:mu-for-infinitessimal}; $\bm{\epsilon}_j$ is deduced
with Equation~\eqref{eq:epsilon_definition}; and the field is
updated, $\field \to \field+\bm{\epsilon}_j$.

\subsubsection{Step choice for a practical
  algorithm} \label{subsubsec:nsteps} When calculating Equation
\eqref{eq:finite-N-steps} as an approximation to Equation
\eqref{eq:matrix-exp-solution}, the accuracy will increase with the
number of steps $N$. One must choose a minimal $N$ (for computational
efficiency) while ensuring that linearly approximating the
modification at each step is sufficiently accurate.

We first perform the calculation with a fixed number of
steps $N_{\text{initial}}$. This gives rise to an initial estimate for
the modified field that we denote $\field_{1,\text{initial}}$. The
error on the resulting modification can be
characterised by the magnitude of $\eta_{\text{initial}}$, where
\begin{equation}
\eta_{\text{initial}} \equiv \field_{1,\text{initial}}^{\dagger} \mat{Q} \,
\field_{1,\text{initial}}^{\vphantom{\dagger}} -q\,.
\end{equation}
Because second-order terms are
neglected in the modification, the error term $\eta_{\text{initial}}$ should scale
inverse-quadratically with the number of steps $N_{\text{initial}}$.
We verified this
behaviour numerically for a variety of fields and modifications.
If $\eta_{\text{initial}}$ is smaller than a desired precision,
$\eta_{\text{target}}$, we retain the initial estimate as our final
output field. Otherwise, the calculation must be repeated; the required number of steps to achieve the
target precision is inferred from the quadratic scaling as
\begin{equation}
N = N_{\text{initial}} \, \sqrt{\frac{\eta_{\text{initial}}}{\eta_{\text{target}}}} ,
\end{equation}
Note that $N_{\text{initial}}$ should be kept small to avoid unnecessary iterations;
$N_{\text{initial}}=10$ has been chosen for our test scenarios below.

\begin{figure*}
    \includegraphics[width=\textwidth]{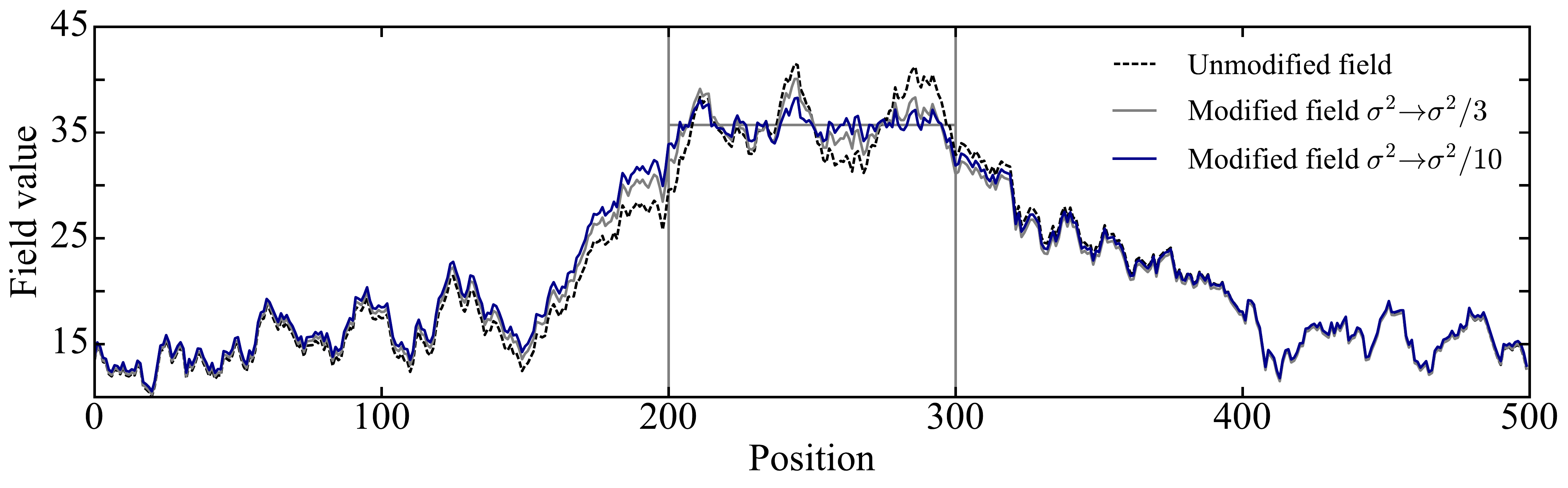}

    \caption{Example genetic modification of a Gaussian random field
      with power spectrum $P(k) \propto (k_0+k)^{-2}$. The unmodified
      and modified fields are shown respectively by dashed and solid
      curves. The region targeted for alteration is enclosed by vertical
      lines. We use simultaneous linear and quadratic modifications to
      conserve the mean value of this region (horizontal line) while
      reducing the small-scale variance by a factor 3 (grey) and 10 (blue).
      In the context of galaxy
      formation, this would maintain the mass of a galaxy and make
      its formation history smoother, while making minimal alterations
      to the large-scale environment.}
 \label{fig:fields}
\end{figure*}

 The final algorithm has a
worst-case complexity of $\mathcal{O}(\eta_{\mathrm{target}}^{-1/2}\,n^3)$, where $n$ is the
number of elements in the field $\field$.  The $n^3$ arises from
matrix multiplications required to compute each step; in
practice the matrices will be sparse either in Fourier space (for the
covariance matrix) or in real space (for the variance $\mat{Q}$
matrix). Therefore, one can speed up the matrix multiplications by
transforming back and forth from real to Fourier space, improving the
complexity to
$\mathcal{O}(\eta_{\mathrm{target}}^{-1/2}\,n \, \log n)$.

The final procedure shares numerous similarities with Newton methods,
used in large-scale optimization (see \citealt{Nocedal2006} for a
comprehensive review). It retains quadratic information in the objective and
linear information in the modification at each step and has a quadratic rate of convergence to the solution.

\subsection{Joint quadratic and linear
  modifications} \label{subsec:linquad} The algorithm above can be
generalised to the case where we have both a quadratic modification
and $p$ linear modifications of the form
$\mat{A} \, \field = \vect{b}$. We first apply the linear
modifications using Equation~\eqref{eq:optim_linear}, then turn to the
iterative quadratic modifications. However if
Equation~\eqref{eq:finite-N-steps} is applied directly, the linear
objective will no longer be satisfied; in other words we need to
enforce $\mat{A} \, \epsilon = \vect{0}$ at each
step. Constructing and solving the appropriate minimisation,
expression~\eqref{eq:epsilon_definition} is replaced by
\begin{equation}
\bm{\epsilon} =
          - \mu \, \mat{C}_0 \, \mat{Q} \, \field
          + \mu \, \mat{C}_0 \, \mat{A}^{\dagger} \,
                  (\mat{A} \, \mat{C}_0  \, \mat{A}^{\dagger})^{-1}
                   \, \mat{A} \,  \mat{C}_0 \, \mat{Q}\, \field
         \, ,
\label{eq:epsilon_linquad}
\end{equation}
where
\begin{equation}
\mu = \frac{1}{2} \, \frac{
       q - \field^{\dagger} \mat{Q} \field
      }
      {\field^{\dagger} \mat{Q} \mat{C}_0 \mat{A}^{\dagger} (\mat{A} \mat{C}_0 \mat{A}^{\dagger})^{-1}
      \mat{A} \mat{C}_0 \mat{Q} \field  -
      \field^{\dagger} \mat{Q} \mat{C}_0 \mat{Q} \field
      } \, .
\end{equation}
These results can be iterated to achieve the final modified
field, in exactly the same way as for the pure-quadratic modification.

Despite the complexity of these expressions, the evaluation will remain
$\mathcal{O} (\eta_{\mathrm{target}}^{-1/2}\,n\,\log n)$ for reasons discussed previously.
To help interpret the method, there is a clear geometric meaning
for each term, which we present in
Appendix~\ref{app:geometry}.

%%%%%%%%%
% Toy model section
%%%%%%%%%
\section{Demonstration} \label{sec:demonstration}

In this Section we demonstrate our algorithm in a $n$-pixel,
one-dimensional setting as a proof of concept and as a reference for
future implementation on cosmological simulations. We choose an
example red power spectrum, as typically encountered on the scales
from which galaxies collapse. Specifically, we adopt
$P(k) = P_0 \, (k_0+k)^{-2}$, where $P_0$ is an arbitrary
normalisation and $k_0 = 2\pi / n$, an offset that prevents divergence
of $P(k)$ at $k=0$.

\subsection{Defining an example modification}

The framework developed in Section \ref{sec:quadratic} can alter any
property that is quadratic in the field by suitable choice of
$\mat{Q}$.  We now specialise to the case that $\mat{Q}$ corresponds
to the variance of a length-$R$ region of the field. We start by defining the
windowing operator $\mat{W}$ as a rectangular matrix picking
out the desired $R$ entries from the $n$ pixels in $\field$. To
calculate the variance of the region, one then calculates
$\field^{\dagger} \mat{Q}_{\sigma^2} \, \field$ where $\mat{Q}_{\sigma^2}$ can be written
\begin{equation}
\mat{Q}_{\sigma^2} = \frac{1}{R^2} \mat{W}^{\dagger} \left( R \, \mat{I} - \vect{1} \otimes \vect{1}
\right) \mat{W}\,.\label{eq:window-variance}
\end{equation}
Here, $\mat{I}$ is the $R \times R$ identity matrix and
$\vect{1}$ is a length-$R$ vector of ones. Expression~\eqref{eq:window-variance}
is readily verified by constructing
$\field^{\dagger} \mat{Q}_{\sigma^2}  \, \field$ and seeing that it does boil down to
the variance of the chosen region.

We wish to consider the field variance only on scales smaller than the
region size (corresponding to substructures with mass lower than that
of the parent halo).  To achieve this, $\mat{Q}_{\sigma^2}$ can be high-pass
filtered; we use a standard Gaussian high-pass filter $\mat{F}$ where
in Fourier space the elements of $\tilde{\mat{F}}$ are given by
\begin{equation}
\tilde{F}_{lm} = \delta_{lm} \left(1 - \exp
\left[-\frac{1}{2}\left(\frac{k_l}{k_f}\right)^2 \right] \right) \, .
\end{equation}
Here, $k_l = 2\pi l/n$ is the wavenumber of the $l$-th Fourier series
element and $k_f$, the filtering scale, is defined in our case by
$k_f = 2\pi / R$. The most appropriate choice of filtering scales and
shapes in the context of cosmological simulations will be discussed in
a forthcoming paper.

In real space the matrix $\mat{F}$ is defined by
$\mat{F} = \mat{U}^{\dagger} \tilde{\mat{F}} \mat{U}$ where $\mat{U}$
is the unitary Fourier transform matrix. Finally, to localise the
target modification fully, we can re-window the matrix after
smoothing. The operator $\mat{W}^{\dagger} \mat{W}$ achieves this by
setting pixels outside the target window to zero.  With this set of
choices, the final quadratic objective is set by
\begin{align}
\mat{Q} & \equiv \mat{W}^{\dagger} \mat{W} \mat{F}^{\dagger}
\mat{Q}_{\sigma^2} \mat{F}  \mat{W}^{\dagger} \mat{W}  \nonumber \\
& = \frac{1}{R^2} \mat{W}^{\dagger} \mat{W} \mat{F}^{\dagger}
 \mat{W}^{\dagger} \left( R \mat{I} - \vect{1} \otimes \vect{1}
\right) \mat{W} \mat{F}  \mat{W}^{\dagger} \mat{W} \, \textrm{.}
\label{eq:full-constraint-matrix}
\end{align}
In practice, we never calculate the matrix $\mat{Q}$ explicitly but
rather implement a routine to efficiently calculate $\mat{Q} \, \field$
for any field $\field$, which is then used by the algorithm described
in Section~\ref{sec:quadratic}. The ability to bypass storing or
manipulating $\mat{Q}$ is essential to permit the computation to operate on
a 3D cosmological simulation.

\subsection{Results}

Figure~\ref{fig:fields} shows examples of modified fields obtained
with our algorithm. We alter the variance of a region of width
$R = 100$ pixels enclosed by vertical lines, showing two quadratic
modifications with the variance reduced by a factor 3 (light grey) and
a factor 10 (dark blue). In both cases, the mean of the field is held
fixed at the unmodified value (horizontal line). In the setting of a
cosmological simulation, we expect to be able to fix the parent halo
mass (through the mean value) while modifying the smoothness of
accretion (through the variance).

We verified that these fields achieve the linear modification
$\mat{A} \, \field_1 - \vect{b}$ to within numerical accuracy and the
quadratic modification $\field_1^{\dagger} \mat{Q} \field_1 - q$ to
$\eta_{\text{target}} = 10^{-6}$ accuracy. The heights of small-scale
peaks inside the enclosed region are successfully reduced and brought
closer to the mean value for the modified fields.  Visually, it can be
seen that the changes to the field are minimal, maintaining as much as
possible of the structure of the unmodified field in the modified
versions. This underlines how the analytic minimisation,
Equation~\eqref{eq:optim_quadratic}, and its refinement to a
linearised procedure (Section~\ref{subsec:linearised method}) agrees
with the intuitive sense of making minimal changes.  The different
versions of the field form a continuous family as illustrated by the
smooth deformation when reducing the variance by different factors.

Despite the modification objective $\mat{Q}$ being strictly confined
to the target region, modifications can be seen to ``leak'' outside
(beyond the vertical lines). This
effect, which is also seen in linear GMs, is an intentional aspect of
the minimisation construction -- any sharp discontinuities in
the field value or its gradients would give rise to a power spectrum
inconsistent with the ensemble.  In this specific example, the leakage appears
more significant to the left than to the right of the target
region. In general, the algorithm is spatially symmetric but its
effect in any given case is not.

\begin{figure}
    \centering
    \includegraphics[width=\columnwidth]{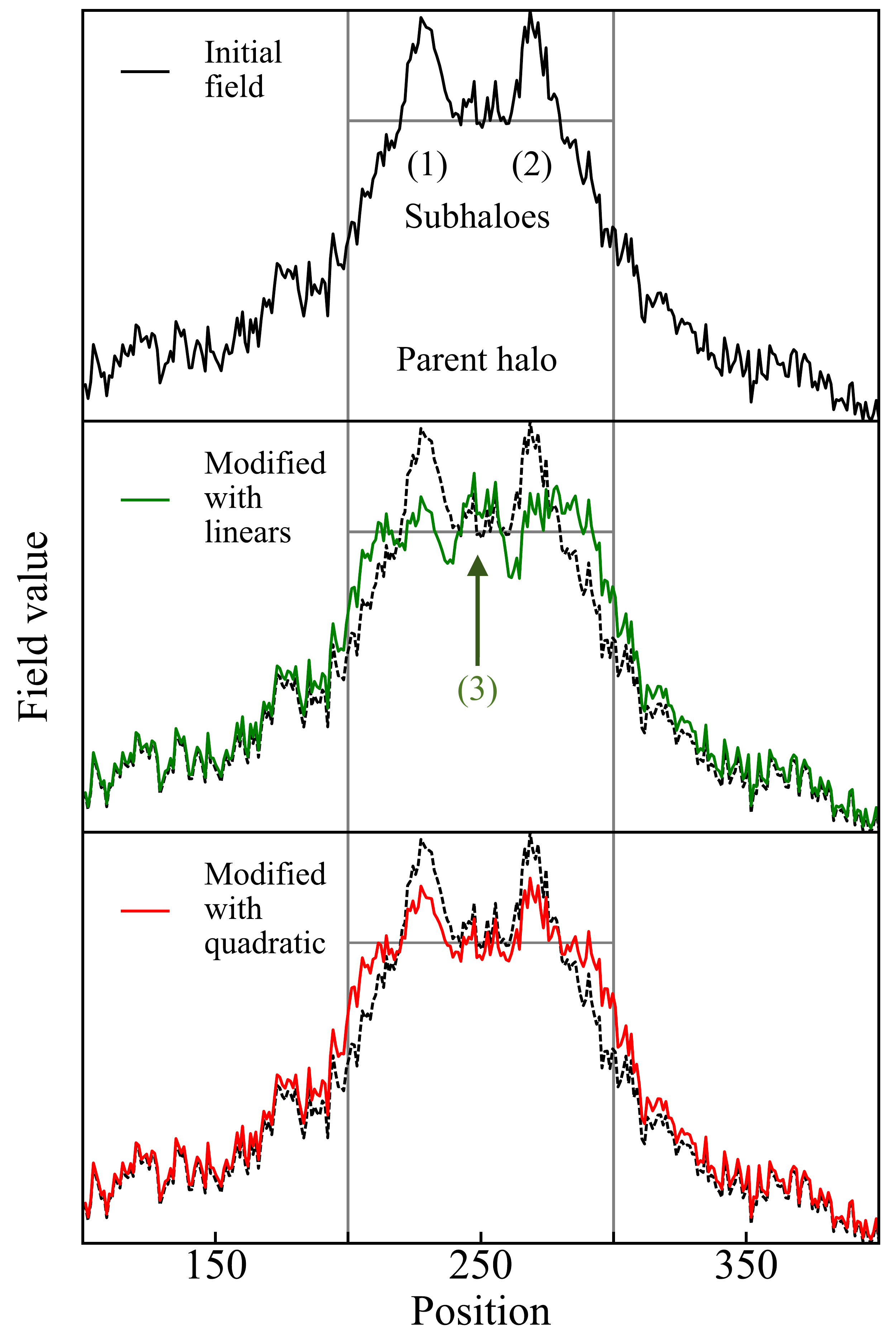}

    \caption{Comparison of pure linear against combined
      linear-quadratic GMs. \textit{ Top panel:} The unmodified field
      contains three distinct features: a broad overdensity that would
      generate a parent halo (enclosed by vertical lines) as well as
      two localised substructures labelled (1) and (2) that would lead
      to a merger during the formation history. The objective is to
      reduce the peak heights of these substructures while conserving
      the mean height of the parent (horizontal line). \textit{Middle
        panel:} a GM field with linear modifications designed to
      bring peaks (1) and (2) to the mean value of the broad
      region. This approach has successfully smoothed the peak
      structure. However, as explained in the text, it suffers from
      the creation of an artificial substructure~(3).  \textit{ Bottom
        panel:} The same objective has been achieved through a
      variance modification. This quadratic modification does not
      require identifying individual subhaloes and
      by construction prevents unhelpful compensations such as~(3).}
 \label{fig:comparison}
\end{figure}

\section{Discussion} \label{sec:discussion}
\subsection{The advantage of quadratic over linear modifications}
\citet{Pontzen2017} showed that using multiple linear modifications was
sufficient to change the merger ratios in the history of a galaxy;
substructures can be diminished or enhanced by manually modifying individual peak heights.

Nonetheless, we expect the new quadratic approach to bring
considerable benefits when making such manipulations; the advantages are
illustrated in Figure~\ref{fig:comparison}.  The top panel shows a
field representing the density in initial conditions expected for a
halo. The field has a broad overdensity enclosed by vertical lines and
two narrower peaks labelled (1) and (2).  According to the excursion-set formalism
\citep{Bond1991}, (1) and (2) will collapse to form two
separate haloes that later merge. This, together
with smooth accretion, will form the final halo.

Suppose we wish to generate a smoother accretion history by reducing
the heights of peaks (1) and (2) while maintaining the large-scale
overdensity. In the original approach, we use linear GMs to set the
mean values of the peaks to the mean value of the broad overdensity
(horizontal line).  The middle panel of Figure~\ref{fig:comparison}
presents the resulting field.  However, a number of problems arise when
performing the alteration using this approach.
\begin{enumerate}
  \item We had to identify (1) and (2) as
        the most interesting substructures and define specific
        modifications for each. In the context of N-body simulations, this requires manually
        identifying which particles of the initial conditions constitute each individual subhalo.
  \item More importantly,
        spatially neighbouring modifications interact and create
        new substructures (peak labelled (3) in our example). One solution
        to prevent the appearance of new substructures could be to
        add a new linear objective forcing problematic regions such as (3) to remain unmodified.
        Identifying and mitigating side effects in this way adds a layer of complexity
        to the linear GM procedure. Depending on the specific problem and the number
        of modifications at play, time spent at this tuning phase can rise steeply.
\end{enumerate}

On the other hand, a single quadratic modification can avoid these
problems by defining a variance target across the region. The third
panel of Figure~\ref{fig:comparison} shows the same field with
variance reduced by a factor 10 (using the method from
Section~\ref{sec:demonstration}).  The two local peaks are
successfully reduced in amplitude while conserving the remaining
small-scale structure of the parent halo.  By construction, the
variance modification naturally avoids compensation problems inherent
to linear GMs.  For this reason, quadratic GMs provide a cleaner,
streamlined way to control merger histories.

\subsection{Multiple quadratic modifications}
The formalism discussed so far applies a single
quadratic modification to a field (possibly in combination with linear
objectives). Simultaneously applying
\textit{multiple} quadratic modifications would allow one to act
concurrently on two separate haloes, or to further fine-tune the merger
history of a single object. For instance, decreasing the variance on intermediate scales
while increasing on small scales should increase the frequency of
minor mergers.

To study this generalisation, we introduce $i=1,\cdots,P$ quadratic modifications, each with matrix $\mat{Q}_i$.
For an infinitesimal update, the change in the field $\bm{\epsilon}$ is then given by
\begin{align}
\label{eq:step-multiple}
&\bm{\epsilon} = \sum_i \mu_{i} \mat{C}_0 \mat{Q}_i \field \, , \\
&\text{with} \quad \field^{\dagger} \mat{Q}_i \field +
                2 \field^{\dagger} \mat{Q}_i \bm{\epsilon} = q_i \quad \text{for all} \ i,
                \label{eq:modif-multiple}
\end{align}
where $\mu_{i}$ are the Lagrange multipliers associated with each
modification.  Equation~\eqref{eq:modif-multiple} defines a system of
$P$ equations to be solved. The resulting value of a specific $\mu_i$
depends on the whole set of $q_i$ and $\mat{Q}_i$, i.e. modifications
are interdependent.

In the same way as Section~\ref{subsec:linearised method}, the update~\eqref{eq:step-multiple}
can be iterated to create finite changes. Performing $N$ steps, the modified field reads
\begin{equation}
\field_1 = \prod_{j=0}^{N} \left( \mat{I} + \sum_{i=0}^{P}\mu_{ij} \,
  \mat{C}_0 \, \mat{Q}_i \right) \field_0\, ,
  \label{eq:multiple-quad-iteration}
\end{equation}
where $\mu_{ij}$ is the multiplier $\mu_{i}$ at step $j$.
However in the limit that the number of steps $N \to \infty$,
convergence to the matrix exponential,
\begin{equation}
\field_1 = \exp \left( \sum_i  \alpha_i \mat{C}_0 \mat{Q}_i \right)  \field_0 \, ,
  \label{eq:multiple-quad-expo}
\end{equation}
is only guaranteed if either the $\mat{Q}_i$ commute with respect to
$\mat{C}_0$ (i.e.
$\mat{Q}_i \mat{C}_0 \mat{Q}_j = \mat{Q}_j \mat{C}_0 \mat{Q}_i$) or
each $\mu_{ij}$ is directly proportional to $\alpha_i$. Because
$\alpha_i$s are not known in advance, the latter option is hard to
arrange; the previously noted interdependence of the $\mu_i$s on all
$q_i$ and $\mat{Q}_i$ exacerbates the difficulty.

With our current algorithms, convergence to the matrix exponential is
therefore only assured when the $\mat{Q}_i$ matrices commute. The
easiest way to arrange for the commutation is to use orthogonal
modifications, i.e.
\begin{equation}
\mat{Q}_i \mat{C}_0 \mat{Q}_j \approx 0 \, .
\label{eq:orthogonality}
\end{equation}
Physically, this requirement can be achieved by imposing modifications
that are spatially separated by a sufficient number of correlation
lengths or address distinct Fourier modes. This condition even allows
one to apply the formalism of Section~\ref{subsec:linearised method} to
each modification one-by-one and still converge to the correct overall
matrix exponential of Equation~\eqref{eq:multiple-quad-expo}. We leave the case
of non-orthogonal multiple quadratic modifications to further work.

\section{Conclusions}\label{sec:concl}

We have presented an efficient algorithm to modify the variance in a
particular region of a Gaussian random field realisation, with the aim
of manipulating initial conditions for cosmological simulations. The
modification produces a field that is as close as possible
to the original realisation. In this way it
provides a route for controlled tests of galaxy formation where
multiple versions of the same galaxy are simulated within a fixed
cosmological environment, but with altered accretion history.

We
argued that quadratic controls, as developed here, offer a useful
complement to the existing linear technique \citep{Roth2016}.
In particular, variance on different filtering scales relates to dark
matter halo substructure and merging history
(\citealt{Press1974, Bond1991}). The new algorithm can construct GM
fields with simultaneous control on the mean value and filtered
variance of a region (Figure~\ref{fig:fields}).  This provides a
route to altering merger history and accretion over the lifetime of a given
halo in a way that is more streamlined than modifying individual
substructures (see Figure~\ref{fig:comparison}).

In both linear and quadratic GM, the algorithm searches for fields
which are nearby in the sense of the $\chi^2$ distance measure.  In
the quadratic case, this definition is further refined: for
large shifts in the control parameter $q$ (which represents the
variance in our test cases), the path through field space is defined
by following a series of small shifts. Each of these individually minimize the
$\chi^2$ distance travelled. We demonstrated a formal convergence
property for this series and argued that the approach is desirable for
(a) returning a continuously-deforming field $\field$ as a function of
the changing target variance $q$; (b) being reversible, so that returning
the variance to its initial value also returns the field to its
initial state; (c) being numerically tractable even for 3D zoom
simulations.

In the process, we clarified the mathematical formulation of GM,
carefully distinguishing it from the constrained ensemble of
\citetalias{HR1991} (see Figure~\ref{fig:overview} for an overview).
The status of fields constructed in the two approaches is distinct --
unlike constrained realisations, GMs should be seen as a mapping
between two fields from the same ensemble. In the case of quadratic
objectives such as variance, even the cosmetic similarities between
constraints and modifications are lost (Appendix
\ref{app:sampling_quadratic}).

The next step is an implementation of the new algorithm in a full
$N$-body initial conditions generator, including on varying-resolution
grids appropriate to zoom simulations. This will be presented in a
forthcoming paper where we will evaluate the effectiveness of
quadratic GM (alongside the existing linear technique) for
constructing controlled tests of galaxy formation.

\section*{Acknowledgements}
We thank Am\'elie Saintonge for helpful discussions and Hiranya Peiris
for useful comments on a draft manuscript.  MR acknowledges supports
from the Perren Fund and the IMPACT fund.  AP is supported by the
Royal Society.

%%%%%%%%%%%%%%%%%%%%%%%%%%%%%%%%%%%%%%%%%%%%%%%%%%


\bibliographystyle{mnras}
\bibliography{Biblio}

\begin{thebibliography}{}
\makeatletter
\relax
\def\mn@urlcharsother{\let\do\@makeother \do\$\do\&\do\#\do\^\do\_\do\%\do\~}
\def\mn@doi{\begingroup\mn@urlcharsother \@ifnextchar [ {\mn@doi@}
  {\mn@doi@[]}}
\def\mn@doi@[#1]#2{\def\@tempa{#1}\ifx\@tempa\@empty \href
  {http://dx.doi.org/#2} {doi:#2}\else \href {http://dx.doi.org/#2} {#1}\fi
  \endgroup}
\def\mn@eprint#1#2{\mn@eprint@#1:#2::\@nil}
\def\mn@eprint@arXiv#1{\href {http://arxiv.org/abs/#1} {{\tt arXiv:#1}}}
\def\mn@eprint@dblp#1{\href {http://dblp.uni-trier.de/rec/bibtex/#1.xml}
  {dblp:#1}}
\def\mn@eprint@#1:#2:#3:#4\@nil{\def\@tempa {#1}\def\@tempb {#2}\def\@tempc
  {#3}\ifx \@tempc \@empty \let \@tempc \@tempb \let \@tempb \@tempa \fi \ifx
  \@tempb \@empty \def\@tempb {arXiv}\fi \@ifundefined
  {mn@eprint@\@tempb}{\@tempb:\@tempc}{\expandafter \expandafter \csname
  mn@eprint@\@tempb\endcsname \expandafter{\@tempc}}}

\bibitem[\protect\citeauthoryear{{Bardeen}, {Bond}, {Kaiser}  \&
  {Szalay}}{{Bardeen} et~al.}{1986}]{BBKS}
{Bardeen} J.~M.,  {Bond} J.~R.,  {Kaiser} N.,   {Szalay} A.~S.,  1986, \mn@doi
  [\apj] {10.1086/164143}, 304, 15

\bibitem[\protect\citeauthoryear{{Bertschinger}}{{Bertschinger}}{1987}]{Bertschinger1987}
{Bertschinger} E.,  1987, \mn@doi [\apjl] {10.1086/185066}, 323, L103

\bibitem[\protect\citeauthoryear{{Bertschinger}}{{Bertschinger}}{2001}]{Bertschinger2001}
{Bertschinger} E.,  2001, \mn@doi [\apjs] {10.1086/322526}, 137, 1

\bibitem[\protect\citeauthoryear{{Bond}, {Cole}, {Efstathiou}  \&
  {Kaiser}}{{Bond} et~al.}{1991}]{Bond1991}
{Bond} J.~R.,  {Cole} S.,  {Efstathiou} G.,   {Kaiser} N.,  1991, \mn@doi
  [\apj] {10.1086/170520}, 379, 440

\bibitem[\protect\citeauthoryear{{Di Matteo}, {Springel}  \& {Hernquist}}{{Di
  Matteo} et~al.}{2005}]{DiMatteo2005}
{Di Matteo} T.,  {Springel} V.,   {Hernquist} L.,  2005, \mn@doi [\nat]
  {10.1038/nature03335}, 433, 604

\bibitem[\protect\citeauthoryear{{Hernquist}}{{Hernquist}}{1993}]{Hernquist1993}
{Hernquist} L.,  1993, \mn@doi [\apjs] {10.1086/191784}, 86, 389

\bibitem[\protect\citeauthoryear{{He{\ss}}, {Kitaura}  \&
  {Gottl{\"o}ber}}{{He{\ss}} et~al.}{2013}]{HeB2013}
{He{\ss}} S.,  {Kitaura} F.-S.,   {Gottl{\"o}ber} S.,  2013, \mn@doi [\mnras]
  {10.1093/mnras/stt1428}, 435, 2065

\bibitem[\protect\citeauthoryear{{Hoffman} \& {Ribak}}{{Hoffman} \&
  {Ribak}}{1991}]{HR1991}
{Hoffman} Y.,  {Ribak} E.,  1991, \mn@doi [\apjl] {10.1086/186160}, 380, L5

\bibitem[\protect\citeauthoryear{{Hoffman}, {Pomar{\`e}de}, {Tully}  \&
  {Courtois}}{{Hoffman} et~al.}{2017}]{Hoffman2017}
{Hoffman} Y.,  {Pomar{\`e}de} D.,  {Tully} R.~B.,   {Courtois} H.~M.,  2017,
  \mn@doi [Nature Astronomy] {10.1038/s41550-016-0036}, 1, 0036

\bibitem[\protect\citeauthoryear{{Hopkins}, {Kere{\v s}}, {Murray}, {Quataert}
  \& {Hernquist}}{{Hopkins} et~al.}{2012}]{Hopkins2012}
{Hopkins} P.~F.,  {Kere{\v s}} D.,  {Murray} N.,  {Quataert} E.,   {Hernquist}
  L.,  2012, \mn@doi [\mnras] {10.1111/j.1365-2966.2012.21981.x}, 427, 968

\bibitem[\protect\citeauthoryear{Nocedal \& Wright}{Nocedal \&
  Wright}{2006}]{Nocedal2006}
Nocedal J.,  Wright S.~J.,  2006, Numerical Optimization, 2nd edn.
Springer

\bibitem[\protect\citeauthoryear{Pontzen, Tremmel, Roth, Peiris, Saintonge,
  Volonteri, Quinn  \& Governato}{Pontzen et~al.}{2017}]{Pontzen2017}
Pontzen A.,  Tremmel M.,  Roth N.,  Peiris H.~V.,  Saintonge A.,  Volonteri M.,
   Quinn T.,   Governato F.,  2017, \mn@doi [\mnras] {10.1093/mnras/stw2627},
  465, 547

\bibitem[\protect\citeauthoryear{{Porciani}}{{Porciani}}{2016}]{Porciani2016}
{Porciani} C.,  2016, \mn@doi [\mnras] {10.1093/mnras/stw2222}, 463, 4068

\bibitem[\protect\citeauthoryear{{Press} \& {Schechter}}{{Press} \&
  {Schechter}}{1974}]{Press1974}
{Press} W.~H.,  {Schechter} P.,  1974, \mn@doi [\apj] {10.1086/152650}, 187,
  425

\bibitem[\protect\citeauthoryear{{Roth}, {Pontzen}  \& {Peiris}}{{Roth}
  et~al.}{2016}]{Roth2016}
{Roth} N.,  {Pontzen} A.,   {Peiris} H.~V.,  2016, \mn@doi [\mnras]
  {10.1093/mnras/stv2375}, 455, 974

\bibitem[\protect\citeauthoryear{{Sorce} et~al.,}{{Sorce}
  et~al.}{2016}]{Sorce2016a}
{Sorce} J.~G.,  et~al., 2016, \mn@doi [\mnras] {10.1093/mnras/stv2407}, 455,
  2078

\bibitem[\protect\citeauthoryear{{Wang} et~al.,}{{Wang}
  et~al.}{2016}]{Wang2016}
{Wang} H.,  et~al., 2016, \mn@doi [\apj] {10.3847/0004-637X/831/2/164}, 831,
  164

\makeatother
\end{thebibliography}

%%%%%%%%%%%%%%%%%%%%%%%%%%%%%%%%%%%%%%%%%%%%%%%%%%

%%%%%%%%%%%%%%%%% APPENDICES %%%%%%%%%%%%%%%%%%%%%

\appendix
\section{Constructing constrained ensembles for quadratic
  constraints} \label{app:sampling_quadratic}

In Section~\ref{sec:reformulation}, we contrasted the notion of a
linearly-constrained ensemble (Section~\ref{subsec:sampling_linear})
against that of genetic modifications (Section~\ref{subsec:optimization_linear}). While conceptually
different, constrained ensembles can be sampled using the
\citetalias{HR1991} procedure which can in turn be seen as applying suitable
modifications to realisations from the unconstrained pdf.

In this Appendix we show that there is no such similarity between
quadratically-constrained ensembles and quadratically modified
fields. To put it another way, there is no \citetalias{HR1991}-like
method for generating samples from a quadratically-constrained
ensemble.

Following the same Bayesian argument as in
Section~\ref{subsec:sampling_linear}, we define the
quadratically-constrained ensemble for a fixed $\mat{Q}$ and $q$ by
\begin{equation}
P(\field | q) \propto \exp(-\frac{1}{2} \field^{\dagger} \mat{C}_0^{-1} \field) \
              \delta_D(\field^{\dagger} \, \mat{Q} \, \field - q) \, .
              \label{app:eq:true-quadratic-const-ensemble}
\end{equation}
We will show that the modification procedure does not generate samples
from the ensemble~\eqref{app:eq:true-quadratic-const-ensemble}, even when
$\mat{Q}$ and $q$ are known and fixed in advance.

We start by defining the alternative ensemble $P(\field_1 | q)$ to be
that sampled by drawing an unconstrained field from $P(\field)$ and
using the GM procedure to enforce the constraint
$\field^{\dagger} \mat{Q} \field = q$. In
Section~\ref{subsec:linearised method}, the mapping
$\field_0 \to \field_1$ was given by
\begin{equation}
\label{app:eq:expon}
\field_1 = \exp \left(\alpha(\field_0,q) \, \mat{C}_0 \, \mat{Q} \right) \field_0 \, ,
\end{equation}
where $\mat{C}_0$ is the covariance matrix of the Gaussian
distribution $P(\field)$ and the value of $\alpha(\field_0,q)$ is
implicitly defined by the need to satisfy the quadratic constraint
$\field_1^{\dagger} \, \mat{Q} \, \field_1= q$.

To incorporate this implicit requirement to choose the correct value
of $\alpha$ into an expression for the ensemble, we make use of Bayes'
theorem:
\begin{gather}
  \begin{split}
  \label{app:eq:ensemble}
P(\field_1 | q) &=  \int P(\field_1 | \alpha) P(\alpha | q) \,\dd\alpha \, \\
                &= \iint P(\field_1 | \alpha, \field_0) P(\alpha | q,
                \field_0) P(\field_0) \,\dd\alpha\, \dd\field_0 \, \\
 & = \iint P(\field_1 | \alpha, \field_0) P(q |\alpha, \field_0)
                  \frac{P(\alpha| \field_0)}{P(q)} P(\field_0)\, \dd\alpha
                  \, \dd\field_0 \, .
\end{split}
\end{gather}
Note that the constraint demands $P(q|\alpha,\field_0) =
\delta_D(\field_1^{\dagger} \mat{Q} \field_1 - q)$, where $\field_1$
and $\field_0$ are related by the condition~\eqref{app:eq:expon}.
Writing out the
normalisation condition for $P(\alpha|q,\field)$ then gives
\begin{align}
1 & = \int P(\alpha|q,\field)\, \dd \alpha = \int
\delta_D(\field_1^{\dagger} \mat{Q} \field_1 - q) \frac{P(\alpha |
    \field_0)}{P(q)}\,\dd \alpha\, .
\end{align}
Because $\mat{Q}$ and $\mat{C}_0 \mat{Q}$ are positive semi-definite,
$q$ is a monotonically increasing function of $\alpha$; there is only
one value of $\alpha$ which satisfies the Dirac delta function on the
right-hand-side. Consequently, we can perform the integration by a
change of variables to yield
\begin{align}
\left.  \frac{P(\alpha |
    \field_0)}{P(q)} \right|_{\field_1^{\dagger} \mat{Q} \field_1 = q}  =
\left. \frac{\partial }{\partial \alpha}\right|_{\field_0} \left( \field_1^{\dagger} \mat{Q}
  \field_1 \right) = 2 \field_1^{\dagger} \mat{QC}_0 \mat{Q} \field_1\,.
\end{align}

Substituting this result back into Equation~\eqref{app:eq:ensemble} and
performing the integral over $\field_0$ using
$P(\field_1 | \alpha, \field_0) = \delta_{D}(\field_1 - \exp{(\alpha \, \mat{C}_0 \, \mat{Q})} \field_0)$,
one obtains
\begin{multline}
  P(\field_1 | q) \propto  \delta_{D}(q - \field_1^{\dagger} \mat{Q}
  \field_1) \ \field_1^{\dagger} \mat{QC}_0 \mat{Q} \field_1 \\
 \times \int \dd \alpha \left| e^{-\alpha \mat{C}_0 \mat{Q}}
  \right|  \exp\left(-\frac{1}{2} \field_1^{\dagger} e^{-\alpha
    \mat{Q} \mat{C}_0} \mat{C}_0^{-1} e^{-\alpha
    \mat{C}_0 \mat{Q}} \field_1 \right) \, ,
\label{app:eq:sampled-quadratic-ensemble}
\end{multline}
where normalisation factors depending only on $\mat{C}_0$ have been
dropped. This expression no longer has any explicit reference to
$\field_0$, which was our primary aim. It can now be compared with the
distribution for a true constrained ensemble,
Equation~\eqref{app:eq:true-quadratic-const-ensemble}.  The two
distributions appear different (as expected given our claim of
inequivalence), but the comparison is complicated by the unsolved
integral over $\alpha$ which obscures the content of the expression.

\begin{figure*}
  \centering
  \begin{minipage}{0.49\textwidth}
    \includegraphics[width=0.9\textwidth]{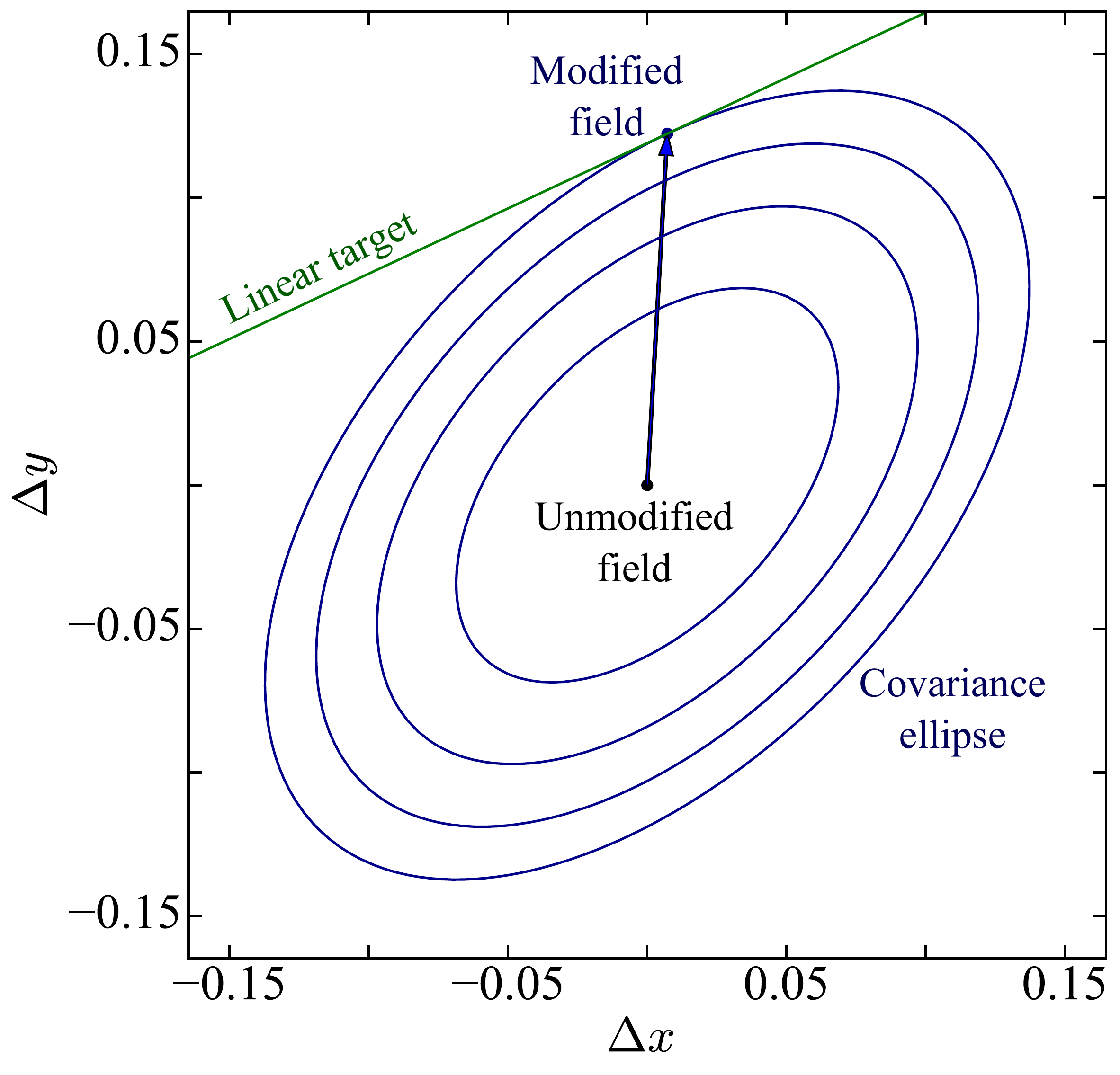}
  \end{minipage}
  \begin{minipage}{0.49\textwidth}
    \includegraphics[width=0.9\textwidth]{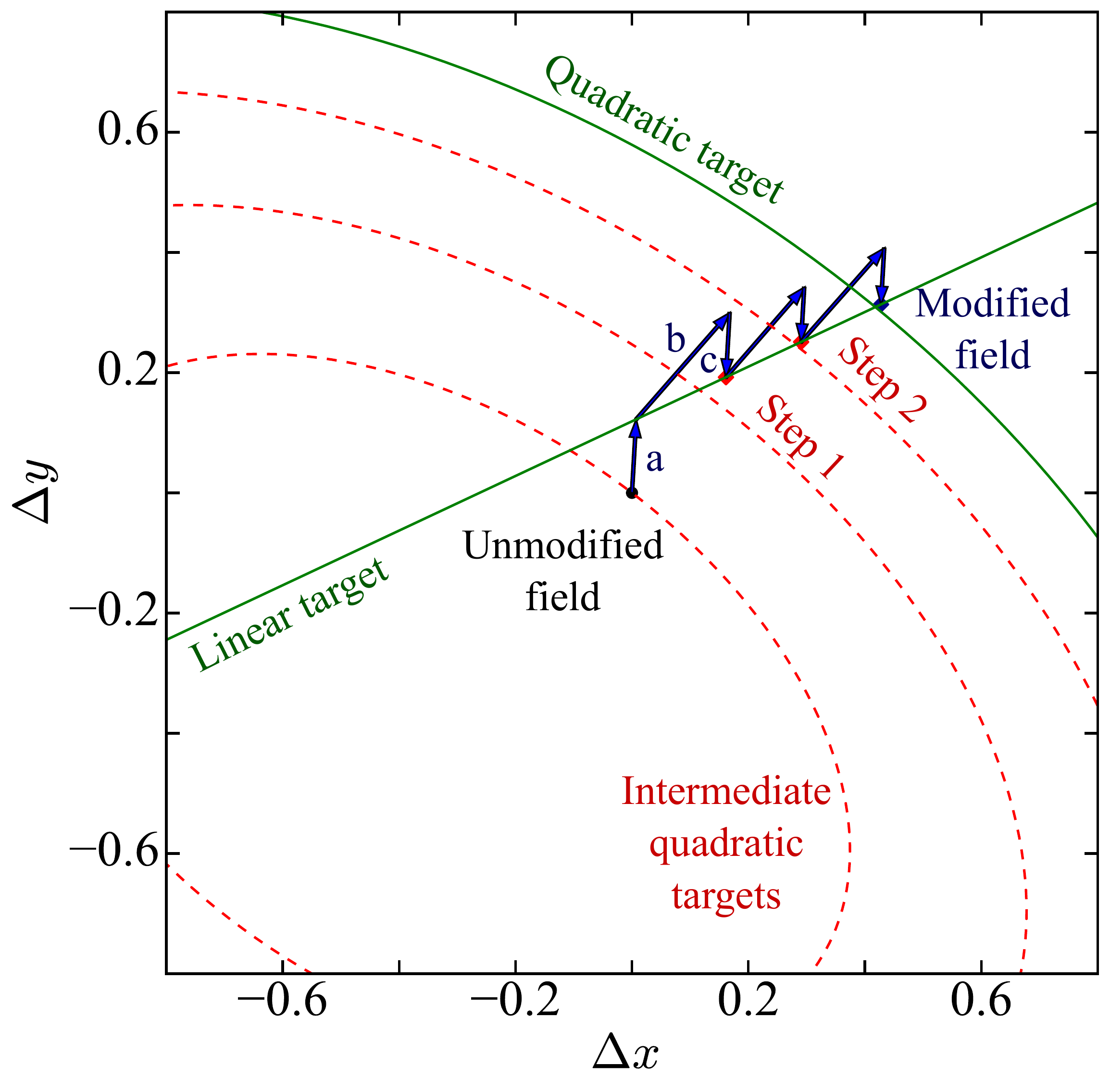}
  \end{minipage}

  \caption{\textit{Left panel}:
    Geometry of linear GM for a field with two components
    $\field=(x,y)^\top$.  The axes represent displacements
    $\Delta \field = (\Delta x, \Delta y)^\top$ from the unmodified
    realisation. The distance measure, Equation~\eqref{eq:distance-measure-2d},
    gives rise to elliptical surfaces
    of constant distance (blue). The linear target corresponds to a
    line (green).  The GM algorithm (arrow) takes the unmodified
    realisation (black dot) to the first intersection between this line
    and ellipses of increasing distance, defining the modified field.
    \textit{Right panel}: Geometry of making simultaneous quadratic and linear
    modifications using the algorithm from Section~\ref{subsec:linearised method}.
    Two target modifications are
    shown, a linear (green line) and a quadratic (green ellipse). The
    algorithm defines intermediate quadratic modifications (red dotted
    ellipses) to step towards the final result. The first operation is the projection
    of the unmodified field onto the linear modification (a); each iterative step then displaces the field
    along the normal of the ellipse (b), and projects it again onto the linear modification (c).}
 \label{fig:ellipses}
\end{figure*}

We can see that this integral will never regenerate the true quadratic
constrained ensemble by taking an illustrative example. Consider a
three-dimensional field $\field_1 = (x,y,z)$ with unit power spectrum
($\mat{C}_0 =\mat{I}$). Let us further choose an explicit form for
$\mat{Q}$ such that
\begin{equation}
\mat{Q} = \left( \begin{array}{ccc}
  0 & 0 & 0 \\
  0 & -1 & 0 \\
  0 & 0 & 1
  \end{array} \right) \Rightarrow
  \mat{e}^{\alpha \mat{C}_0 \mat{Q}} = \left( \begin{array}{ccc}
  1 & 0 & 0 \\
  0 & e^{-\alpha} & 0 \\
  0 & 0 & e^{\alpha}
  \end{array} \right) \, .
\end{equation}
Inserting these results into Equation~\eqref{app:eq:sampled-quadratic-ensemble} gives
\begin{align}
  P(\field_1 | q) &\propto \delta_{D}(q + y^2 - z^2) (y^2 + z^2) \\ \nonumber
   &\times \int_0^{\infty} \frac{\dd \beta}{\beta}
  \exp\left(-\frac{1}{2} \left(x^2 + \beta^{-2} y^2 + \beta^{2} z^2 \right) \right) \,  ,
\end{align}
where we have made the substitution $\beta = e^{-\alpha}$. The integral
over $\beta$ has an analytical solution using the further substitution
$t = (\beta z)^2/2$ and introducing the modified Bessel function
of the second kind
\begin{equation}
K_0(x) = \frac{1}{2}\int_0^{\infty} \frac{\dd t}{t} e^{-t-\frac{x^2}{4}} \, .
\end{equation}
Equation~\eqref{app:eq:sampled-quadratic-ensemble}
can then be evaluated explicitly to obtain
\begin{align}
  P(\field_1 | q)  & \propto e^{-\frac{x^2}{2}} \, \delta_{D}(q + y^2 - z^2) \, (y^2 + z^2)
   \, K_0\left(|yz|\right) \, .
\label{app:eq:example}
\end{align}
For comparison, the quadratic constrained ensemble in this example is given by
\begin{align}
  P(\field | q)  & \propto  e^{-\frac{x^2}{2}} \, \delta_{D}(q + y^2 - z^2) \, e^{-\frac{y^2 +z^2}{2}} \, .
\label{app:eq:quadratic-example}
\end{align}

The distributions defined by~\eqref{app:eq:example} and~\eqref{app:eq:quadratic-example} have
identical $x$-dependence. This is a  general
property: degrees of freedom for which $\mat{Q}$ has a null direction are
unconstrained and, similarly, left
unchanged by our GM transformation. The distribution generated by these degrees of freedom
will therefore coincide at all times with the constrained ensemble.
However, the $y$ and $z$ dependences differ between
Equations~\eqref{app:eq:example} and~\eqref{app:eq:quadratic-example}. In general,
non-null directions in field space will behave
differently between the GM and constrained ensemble cases.

The result establishes that our formulation of quadratic GM as a
matrix exponential mapping does not reproduce a
quadratically-constrained ensemble when used analogously to the
\citetalias{HR1991} algorithm. A similar argument allows one to verify
that applying the alternative non-linear modification specified by
Equation~\eqref{eq:field1-from-mu} also fails to regenerate the
constrained result. In fact, one can go even further and write a
general power series expansion for the mapping between $\field_0$ and
$\field_1$, writing
\begin{equation}
\field_1 = \sum_{i=0}^{\infty} A_i(\mu \mat{C}_0 \mat{Q})^i \field_0\,,
\end{equation}
without further specifying the power series coefficients $A_i$. Even
in this case, which generalises away from a specific mapping, it is not
possible to generate a constrained ensemble from the modification
procedure. This underlines that modifications and constraints need to
be regarded as entirely separate procedures. Only in the linear case
do they appear to be cosmetically related.

\section{Geometrical interpretation} \label{app:geometry}

Throughout the main text, we used fields sampled at a finite number of
points $n$; the resulting algorithms can therefore be interpreted geometrically
as acting on vectors in an $n$-dimensional space. For instance,
\citet{Roth2016} noted that the linear GM procedure is equivalent to an
orthonormal projection of the unmodified field onto a subspace
defining the modification objective (see their appendix A).  In this
Appendix we provide the geometric interpretation for our extended
formulation of GM.

For the purposes of visualising the connection, we use fields with
only two samples, $\field = (x,y)$.  The arguments of this Appendix
generalise to higher dimensions but are easiest to visualise with
$n=2$.  Figure~\ref{fig:ellipses} shows the resulting two-dimensional
geometry in terms of the displacements $\Delta x$ and $\Delta y$ from
the unmodified field. By construction, the unmodified field is
 at the origin.

The left panel shows the elliptical geometry generated by the
covariance matrix in the $\Delta x$ -- $\Delta y$ plane; specifically, the ellipses
are of constant distance $\Delta s^2$ from the origin, where
\begin{equation}
\Delta s^2 \equiv \left(\begin{array}{cc} \Delta x & \Delta y \end{array} \right)
                           \mat{C}_0^{-1}
                   \left(\begin{array}{c} \Delta x \\\Delta y \end{array} \right)
                   \,.
\label{eq:distance-measure-2d}
\end{equation}
The linear objective $\mat{A} \field = \vect{b}$ defines a line in two dimensions.
The modification consists of finding the value of $(\Delta x, \Delta y)$ lying
on the line, while minimising $\Delta s^2$.
Since $\Delta s^2$ is measured in terms of
$\mat{C}_0^{-1}$, the solution does not correspond to the
closest point on the page but to the point at
which a covariance ellipse is tangent to the modification line.

Similarly, the quadratic modifications (right panel of
Figure~\ref{fig:ellipses}) are associated with ellipses of constant
$q = (x,y) \, \mat{Q}\, (x,y)^{\top}$.  These targets are shown as dotted lines;
note that they are centred on $(x,y) = (0,0)$ and therefore appear
offset from the origin in the $\Delta x$ -- $\Delta y$ plane.

The right panel of Figure~\ref{fig:ellipses} also illustrates the
algorithm for finding the modified field with a simultaneous quadratic
and linear objective. For visual clarity, an unrealistically small
($N=3$) number of steps are taken.  We start by defining three
intermediate ellipses (red-dotted) between the value of the
modification at the unmodified field and the target. As explained
in Section~\ref{subsec:linquad}, we first apply the global linear modifications from
Equation~\eqref{eq:optim_linear}
\begin{equation}
\field \to \field - \underbrace{ \mat{C}_0 \, \mat{A}^{\dagger} \,
        (\mat{A} \, \mat{C}_0  \, \mat{A}^{\dagger})^{-1} \, (\mat{A} \, \field - \vect{b} )}_{\text{a}} \, .
\end{equation}
The algorithm then iterates the step
$\bm{\epsilon}$ defined by Equation~\eqref{eq:epsilon_linquad}
\begin{equation}
\bm{\epsilon} =
          - \underbrace{ \mu \, \mat{C}_0 \, \mat{Q} \, \field}_{\text{b}}
          + \underbrace{ \mu \, \mat{C}_0 \, \mat{A}^{\dagger} \,
                  (\mat{A} \, \mat{C}_0  \, \mat{A}^{\dagger})^{-1}
                   \, \mat{A} \,  \mat{C}_0 \, \mat{Q}\, \field}_{\text{c}} \, .
\end{equation}
These operations can be understood geometrically as:
\begin{description}
  \item (a) A projection of the current field on the linear modification. This term is
        similar to the case with linear modifications only.
  \item (b) A displacement along the normal of the ellipse at
        the current field value. This term is
        towards the next intermediate ellipse.
  \item (c) The projection of the previous term back onto the linear modification to ensure that
        both are always satisfied.
\end{description}

Term (c) ensures that the current field at the end of each step always lies on
the linear constraint. Term (a) therefore vanishes after the first step;
it is an overall offset that needs to be
applied only once. Together, (b) and (c) are locally
orthogonalizing the quadratic modification with respect to the global
linear modification. The orthogonalisation must be repeated at each
step since the local linearisation changes as we progress towards the
final value of $q$.

%%%%%%%%%%%%%%%%%%%%%%%%%%%%%%%%%%%%%%%%%%%%%%%%%%

\bsp	% typesetting comment
\label{lastpage}
\end{document}